\documentclass[aps,pra,amssymb,twocolumn,amsmath,superscriptaddress,showpacs,10pt]{revtex4-2}

\usepackage{graphicx}
\usepackage{dcolumn}
\usepackage{bm}
\usepackage{color}
\usepackage{epstopdf}
\usepackage{booktabs}
\usepackage[center]{subfigure}

\usepackage[normalem]{ulem}

\def\br{\bm{\rho}}

\definecolor{darkcyan}{rgb}{0.0, 0.55, 0.55}

\begin{document}

\title{Light-field microscopy with correlated beams for extended volumetric imaging at the diffraction limit}

\author{Gianlorenzo Massaro}
\affiliation{Dipartimento Interuniversitario di Fisica, Universit\`a degli studi di Bari, I-70126 Bari, Italy}
\affiliation{INFN, Sezione di Bari, I-70126 Bari, Italy}

\author{Davide Giannella} 
\affiliation{Dipartimento Interuniversitario di Fisica, Universit\`a degli studi di Bari, I-70126 Bari, Italy}
\affiliation{INFN, Sezione di Bari, I-70126 Bari, Italy}

\author{Alessio Scagliola} 
\affiliation{Dipartimento Interuniversitario di Fisica, Universit\`a degli studi di Bari, I-70126 Bari, Italy}
\affiliation{INFN, Sezione di Bari, I-70126 Bari, Italy}

\author{Francesco Di Lena} 
\affiliation{INFN, Sezione di Bari, I-70126 Bari, Italy}

\author{Giuliano Scarcelli} 
\affiliation{Fischell Department of Bioengineering, University of Maryland, College Park MD 20742, United States}

\author{Augusto Garuccio} 
\affiliation{Dipartimento Interuniversitario di Fisica, Universit\`a degli studi di Bari, I-70126 Bari, Italy}
\affiliation{INFN, Sezione di Bari, I-70126 Bari, Italy}

\author{Francesco V. Pepe} 
\affiliation{Dipartimento Interuniversitario di Fisica, Universit\`a degli studi di Bari, I-70126 Bari, Italy}
\affiliation{INFN, Sezione di Bari, I-70126 Bari, Italy}

\author{Milena D'Angelo}
\affiliation{Dipartimento Interuniversitario di Fisica, Universit\`a degli studi di Bari, I-70126 Bari, Italy}
\affiliation{INFN, Sezione di Bari, I-70126 Bari, Italy}

\begin{abstract}
Light-field microscopy represents a promising solution for microscopic volumetric imaging, thanks to its capability to encode information on multiple planes in a single acquisition. This is achieved through its peculiar simultaneous capture of information on light spatial distribution and propagation direction. However, state-of-the-art light-field microscopes suffer from a detrimental loss of spatial resolution compared to standard microscopes. We propose and experimentally demonstrate a light-field microscopy architecture based on light intensity correlation, in which resolution is limited only by diffraction. We demonstrate the effectiveness of our technique in refocusing three-dimensional test targets and biological samples out of the focused plane. We improve the depth of field by a factor 6 with respect to conventional microscopy, at the same resolution, and obtain, from one acquired correlation image, about $130,000$ images, all seen from different perspectives; such multi-perspective images are employed to reconstruct over $40$ planes within a $1 \,\mathrm{mm}^3$ sample with a diffraction-limited resolution voxel of $20 \times 20 \times 30\ \mu\mathrm{m}^3$. 
\end{abstract}


\maketitle

\section{Introduction}

Rapid imaging of three dimensional samples at the diffraction limit is a long-standing and largely unresolved challenge of microscopy \cite{mertz}. Many attempts are being made to address the need for rapid imaging of large volumes, with acquisition speed sufficient to analyze dynamic biological processes. Progresses in this field include depth focal scanning with tunable lenses \cite{oku,mermillod}, light-sheet illumination \cite{huisken}, or multi-focus multiplexing \cite{maurer}, as well as compressive sensing and computational microscopy techniques \cite{waller2015computational}. In this perspective, light-field microscopy is among the most promising techniques. By detecting both the spatial distribution and the propagation direction of light, in a single exposure, light-field imaging has introduced the possibility to refocus out-of-focus parts of three-dimensional samples, in post-processing. The depth of field (DOF) within the imaged volume can thus be extended by stacking refocused planes at different distances \cite{adelson,ng,microscopy1,focused_pleno,plenoptic_review,muenzel}. However, in its traditional implementation, light-field imaging is affected by the fundamental barrier imposed by the resolution versus DOF compromise. In the microscopic domain, this tradeoff is particularly suboptimal because the required high resolution, which is proportional to the numerical aperture (NA) of the imaging system, strongly limits the DOF, making it necessary to perform multiple scanning to characterize a thick sample \cite{minsky}. In microscopy applications where light-field microscopy could offer a solution to the bottlenecks of long acquisition times (typical of scanning approaches) and unbearably large amount of data (typical of multi-focus multiplexing), its widespread application has been stifled by the degraded resolution, far away from the diffraction limit \cite{ng,focused_pleno}. Nevertheless, fostered by the development of image analysis tools and deconvolution algorithms that provide a partial recovery of resolution \cite{dansereau2013decoding,perez2014super,li2016scalable}, ligthfield imaging has shown its potential in neuroscience applications, where it has been employed to analyze firing neurons in large areas \cite{microscopy4}. A miniaturized version of a ligthfield microscope has also been recently employed to enable microscopy in freely moving mice \cite{skocek}.

In this article, we demonstrate a novel method to perform light-field microscopy with \textit{diffraction-limited resolution}, and discuss its experimental realization and testbed applications. The new technique, capable of beating the classical microscopy limits by exploiting the statistical properties of light \cite{lassen2017superresolution,kviatkovsky2020microscopy,simon2010twin,gattomonticone2014beating,samantaray2017realization,altmann2018quantum}, employs the working principle of so called Correlation Plenoptic Imaging (CPI) \cite{cpi_prl,cpi_qmqm,cpi_technologies,cpi_setups,cpi_exp,cpi_snr}, in which plenoptic (or light-field) imaging is performed at the diffraction limit by measuring correlations between intensity fluctuations at two disjoint detectors \cite{pittman1995optical,gatti,laserphys,valencia,scarcelliPRL}. However, previous CPI architectures were limited to bright-field operation relying on mask objects. Here, we design a CPI architecture suitable for different microscopy modalities (fluorescence, polarization, dark-field) that are critical for biological applications.
To this end, the sample is illuminated with the whole light beam from a chaotic light source \cite{cpm_theory}, rather than by just one beam out of a correlated beam pair \cite{cpi_prl,cpi_setups,cpi_exp}. This enables imaging self-emitting, scattering and diffusive samples, as well as performing birefringent imaging, without sacrificing the retrieved correlation. Further advantages of the proposed Correlation Light-field Microscope (CLM) over previous ones are the speed-up of the image acquisition by over one order of magnitude, and the capability of monitoring the sample through conventional (i.e., intensity-based) diffraction-limited microscopy.

\begin{table*}
\centering
\begin{tabular}{l|cccc}
Microscopy & Resolution & DOF at & DOF extension at & Viewpoint \\ 
method & limit & resolution limit & object resolution & multiplicity
\\ \hline \\
Standard & $\Delta x_0$ & $\displaystyle\frac{\Delta x_0}{\mathrm{NA}_0}$ & $\displaystyle\frac{a}{\mathrm{NA}_0}$ & $1$ \\  \\
Light-field & $N_u\Delta x_0$ & $\displaystyle\frac{N_u^2\Delta x_0}{\mathrm{NA}_0}$ & $N_u\displaystyle\frac{a}{\mathrm{NA}_0}$ & $N_u$ \\  \\
CLM & $\Delta x_0$ & $\displaystyle\frac{\Delta x_0}{\mathrm{NA}_0}$ & $\displaystyle\frac{a^2}{\lambda}$ & $\displaystyle\frac{a}{\Delta x_0}$ \\  \\
\hline
\end{tabular}
\caption{Comparison of resolution and depth-of-field (DOF) limits of three microscopy techniques: standard microscopy (with no directional resolution), standard light-field microscopy, and correlation light-field microscopy (CLM). Here, $\Delta x_0=1.22\lambda/\mathrm{NA}_0$, with $\mathrm{NA}_0$ the numerical aperture of the microscope, is the diffraction-limited resolution. $N_u$ is the number of directional resolution cells per side in standard light-field imaging, and $a$ is the size of the smallest details within the sample. The first and second columns represent the resolution limit in the focal plane and the maximum DOF achievable for objects with details at the resolution limit, respectively. The third column indicate the maximum DOF achievable for object with detail size $a>\Delta x_0$ larger than the resolution limit. The last column reports the number of viewpoints per direction. Properties of the light-field microscope are derived by the general features of light-field imaging devices (see \cite{ng2005fourier} for a detailed discussion), while the resolution and DOF limits in CLM are derived from the theoretical analysis reported in Ref.~\cite{cpm_theory}, as well as in Materials and Methods.}\label{tab:comp}
\end{table*}

The comparison reported in Table~\ref{tab:comp} clarifies the expected theoretical improvements offered by CLM, in terms of both resolution and DOF \cite{cpm_theory}, with respect to both standard microscopy and conventional light-field microscopy \cite{ng2005fourier}. The first column highlights the diffraction-limited imaging capability that CPM shares with conventional microscopy; in contrast, the image resolution of conventional light-field imaging is sacrificed by a factor $N_u$, which enables a proportional DOF improvement. The second and third columns of the table report the DOF of the three methods, respectively, for object details at the resolution limit and object details of arbitrary size. In light-field imaging, the latter represents the refocusing range, while the former determines the axial resolution, namely, the ability to refocus details on different transverse planes along the longitudinal direction, within this refocusing range. In conventional light-field microscopy, increasing the refocusing range (i.e., choosing large values of $N_u$) entails a proportional loss of transverse resolution, and an even more detrimental loss of axial resolution (proportional to $N_u^2$); this generally limits $N_u$ to values smaller than $10$. Furthermore, for object details larger than the resolution limit, both  the DOF of standard microscopy and the refocusing range of conventional light-field microscopy scale linearly with the size of the object details; this is due to the ``circle of confusion'' generated by the finite NA of the imaging system. In CLM, instead, the refocusing range scales quadratically with the size of object details, and is only limited by diffraction at the object (see Refs.~\cite{cpi_prl,cpi_exp} for a detailed discussion).
These are key features in ensuring the unique refocusing advantage of CLM. 
The last column of Table~\ref{tab:comp} also indicates that the refocusing range of both light-field microscopes is strictly related with the viewpoint multiplicity, defined as the number of available viewpoints on the objects. 
In sharp contrast with conventional light-field microscopy, the viewpoint multiplicity of CLM is completely decoupled from the resolution on the focused plane, and can be easily made larger than in conventional light-field imaging by even one order of magnitude without affecting the diffraction-limited resolution; this is especially true for imaging systems with a large NA and for refocusing far away from the focused plane. Most important, the DOF extension capability of CLM is independent on the numerical aperture of the imaging systems (see Material and Methods for details); a large NA can thus be chosen for maximizing the volumetric resolution. This is very different from conventional light-field microscopy, where one has to compromise the numerical aperture with  $N_u$ in order to achieve an acceptable compromise of resolution and DOF.

In this paper, CLM is employed to extend the DOF within a range $6$ times larger than in a conventional microscope, while maintaining diffraction-limited resolution. With just one correlation light-field image, our CLM  collects two-dimensional images from about $130,000$ ($=360 \times 360$) diffraction-limited perspectives, which would entail a loss of resolution by a factor of $360$ in a traditional light-field microscope (see second column of Table~\ref{tab:comp}). 
This scanning-free multi-perspective images are employed to refocus over $40$ planes within a large ($1 \times 1 \times 1 \,\mathrm{mm}^3$) three-dimensional sample with a diffraction-limited voxel of $20 \times 20 \times 30 \,\mu\mathrm{m}^3$, as determined by the numerical aperture of the objective lens (see Table~\ref{tab:comp}).

\begin{figure}
\centering
\includegraphics[width=0.48\textwidth]{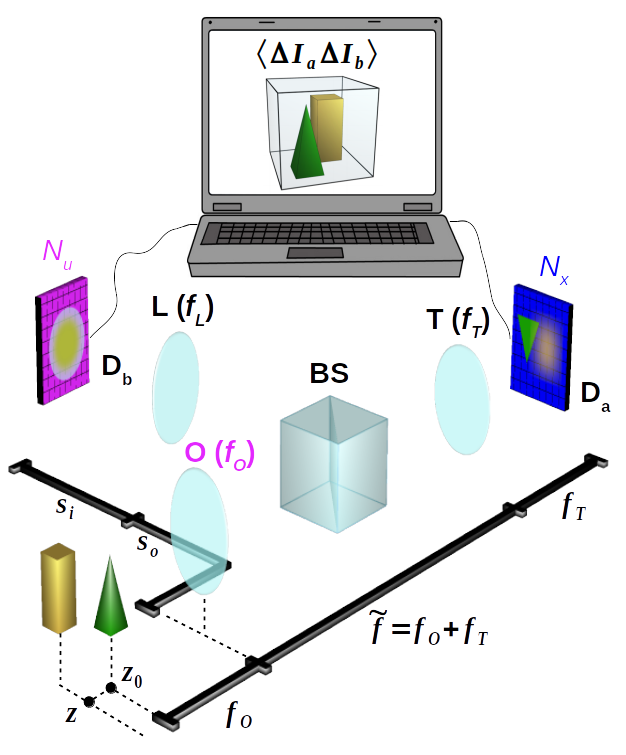}
\caption{Schematic representation of light-field microscopy with correlated beams. Light from the sample (green pyramid and yellow parallelepiped) is split in two optical paths by a beam splitter (BS),  placed after the objective lens (O). The microscope can image only the part of the sample which is at focus (green), while the part outside the DOF is blurred (yellow). Along the transmitted path, the tube lens (T) focuses on detector $\mathrm{D}_a$ (blue) the 3D sample, with a depth of focus defined by the numerical aperture of the objective. Along the reflected path of the BS, the objective lens is imaged on detector $\mathrm{D}_b$ (magenta) by means of an additional lens (L). The intensity patterns registered by the two array detectors, in a collection of $N$ frames, are processed by a computer to reconstruct the correlation function (Eq.\ \eqref{Gammacorr}) encoding three-dimensional plenoptic information on the sample.}\label{fig:plenoptic}
\end{figure}

\section{Results}

\noindent CONCEPT

\hspace{1 cm}

\noindent The Correlation Light-field Microscope, schematically represented in Fig.\ \ref{fig:plenoptic}, is based on a conventional microscope made of an objective lens (O) and a tube lens (T) to reproduce the image of the sample on a high resolution sensor array (detector $\mathrm{D}_a$); this microscope can properly reconstruct only the slice of the three-dimensional object falling within its depth of focus. The capability of CLM to refocus out-of-focus parts of the three-dimensional sample comes from its ability to also gain directional information about light coming from the sample. In our architecture, this is done by means of the beam splitter that reflects a fraction of light emerging from the objective lens toward an additional lens (L), which images the objective lens on a second high resolution sensor array (detector $\mathrm{D}_b$). Further details on the experimental setup are reported in the Material and methods section.

The three-dimensional sample is a chaotic light emitter or, alternatively, a diffusive, transmissive, or reflective sample illuminated by an external chaotic light source. The chaotic nature of light enables light-field imaging thanks to the rich information encoded in \textit{correlations between intensity fluctuations}. In fact, the intensity retrieved by pixels simultaneously illuminated on the two disjoint detectors $\mathrm{D}_a$ and $\mathrm{D}_b$ is employed in CLM to evaluate the correlation function
\begin{equation}\label{Gammacorr}
\Gamma(\bm{\rho}_a,\bm{\rho}_b) = \langle \Delta I_a (\br_a) \Delta I_b (\br_b) \rangle,
\end{equation}
where $\langle\dots\rangle$ is the average over $N$ frames, $I_{a}(\br_{a})$ and $I_{b}(\br_{b})$ are the intensities at the transverse positions $\br_a$ and $\br_b$ on detectors $\mathrm{D}_a$ and $\mathrm{D}_b$, within the same frame, respectively, and $\Delta I_{j} (\br_{j}) = I_{j}(\br_{j}) - \langle I_{j} (\br_{j}) \rangle$, $j=a,b$. 

\begin{figure}
\centering
\includegraphics[width=0.48\textwidth]{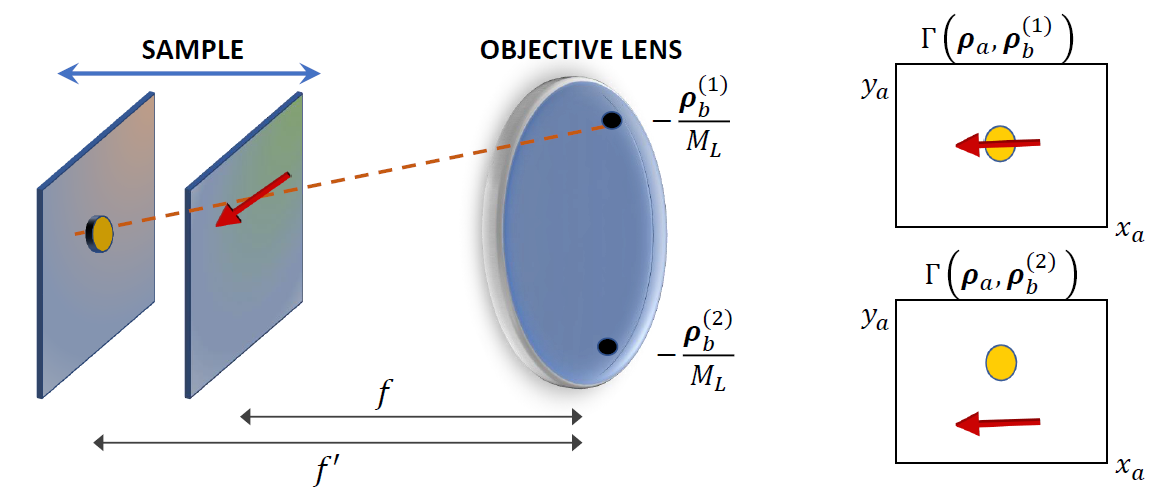}
\caption{Schematic representation of the different points of view on the 3D sample that the correlation function $\Gamma(\br_a,\br_b)$ incorporate (Eq.\ \eqref{geom}); here, $\br_i=(x_i, y_i)$, $i=a,b$. A thick sample is represented through a (yellow) circle and a (red) arrow placed at distances $f$ and $f'$, respectively, from the objective lens. These two details can appear superposed (upper panel on the right) or well separated (lower panel on the right) depending on the specific transverse coordinate $\br_b$ chosen on detector $\mathrm{D}_b$, that corresponds to a point $-\br_b/M_L$ on the objective lens. }\label{fig:views}
\end{figure}

The light-field imaging capability of CLM explicitly emerges when considering the geometrical optics limit of the above correlation function, which reads \cite{cpm_theory}
\begin{equation}\label{geom}
\Gamma(\br_a,\br_b) \sim F^2 \! \left( - \frac{f}{f_T} \br_a - \left( 1 - \frac{f}{f_O} \right) \frac{\br_b}{M_L} \right)
\end{equation}
where $F(\br_s)$ is the intensity profile of light from the sample, $f$ is the distance from the objective of the generic plane within the three-dimensional object, $f_T$ is the focal length of the tube lens, and $M_L$ is the magnification of the image of the objective lens retrieved by $\mathrm{D}_b$. 
When the plane of interest is on focus (i.e., $f=f_O$, with $f_O$ the focal length of the objective), the correlation simply gives a focused image identical to the one retrieved by detector $\mathrm{D}_a$. However, as shown in Fig.\ \ref{fig:views}, points of the three-dimensional samples that are out-of-focus (i.e., they lie in planes at a distance $f \neq f_O$ from the objective) are seen as shifted, and their displacement depends on the specific pixel $\br_b$ chosen on sensor $\mathrm{D}_b$, corresponding to the point $\br_O=-\br_b/M_{L}$ on the objective. In other words, for three-dimensional samples that are thicker than the natural depth of field of the microscope, different values of $\br_b$ correspond to different choices of the point of view on the sample: The correlation function in Eq. \eqref{Gammacorr} has the form of a four-dimensional array, characterized by both detector coordinates $(x_a, y_a, x_b, y_b)$, encoding all the spatial and angular information needed for refocusing and multi-perspective image. By fixing the coordinates $(x_b,y_b)$ of the 4D array, one makes a ``slice'' of the correlation function, which corresponds to selecting an image of the sample from a chosen viewpoint on the objective lens. This property enables to detect the position of sample details, in three dimensions, and to highlight hidden parts of the sample. The refocused image of a sample plane placed at an arbitrary distance $f$ from the objective can be obtained by properly stacking and summing such different perspectives: 
\begin{multline}\label{refocus}
\Sigma_{\mathrm{ref}} (\br_a) = \int \mathrm{d}^2\br_b \Gamma \left( \frac{f_O}{f} \br_a + \left( 1- \frac{f_O}{f} \right) \frac{M}{M_L} \br_b , \br_b \right) \\ 
\sim F^2 \! \left( - \frac{\br_a}{M} \right) 
\end{multline}
with $M=f_T/f_O$ being the natural microscope magnification. The refocusing procedure clearly enables significantly increasing the signal-to-noise ratio with respect to the one characterizing the single perspective associated with $\br_b$. 

Before moving on to the experimental demonstration of CLM, let us spend a few words on the advantages offered by the capability of CLM to collect a large number of viewpoints, as enabled by the large objective plane on which perspectives can be selected, and by the consequent wide angle from which the sample can be observed. First, refocusing shares with 3D imaging the close connection between maximum observation angles and achievable DOF. Second, the higher the number of points of view that are superimposed to obtain the refocused image (see the integration over $\rho_b$ in Eq.~\eqref{refocus}), the more effective will be both the enhancement of features on the plane of interest and the suppression of contributions from neighboring planes, as well known in 3D imaging. 
The total number of perspectives should not be confused with the viewpoint multiplicity shown in Table \ref{tab:comp}. Both in conventional light-field imaging and in CLM, the number of available perspectives depends solely on setup specifications, such as the amount of directions that can be geometrically discriminated, or, in CLM, the resolution characterizing the image the objective lens. The viewpoint multiplicity, on the other hand, depends on the object details size, so that many viewpoints might end up contributing to the same resolution cell as the size of the details is increased. If the multiple perspectives are statistically independent from one another,  their superposition results in a point-wise increase of the signal-to-noise ratio, which is proportional to the squared root of points of view contributing to a given pixel of the final image; the redundancy thus offers a SNR advantage within the final CLM image.

\begin{figure}
\centering
\subfigure[]{\includegraphics[width=0.45\textwidth]{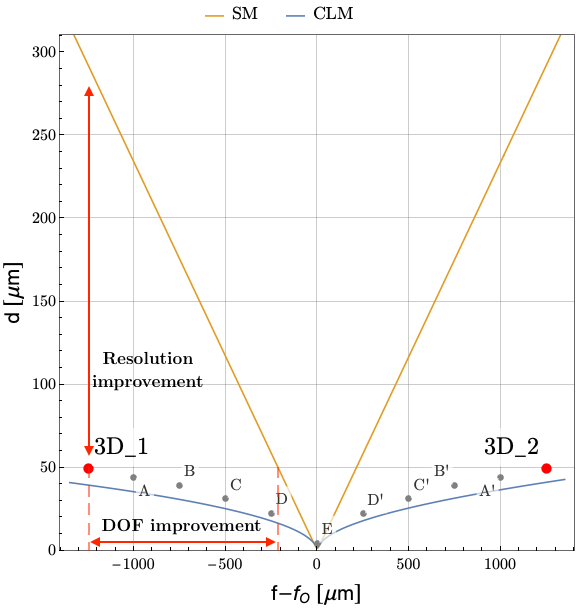}}
\subfigure[\hspace*{75pt}(c)\hspace*{75pt}(d)]{\includegraphics[width=0.48\textwidth]{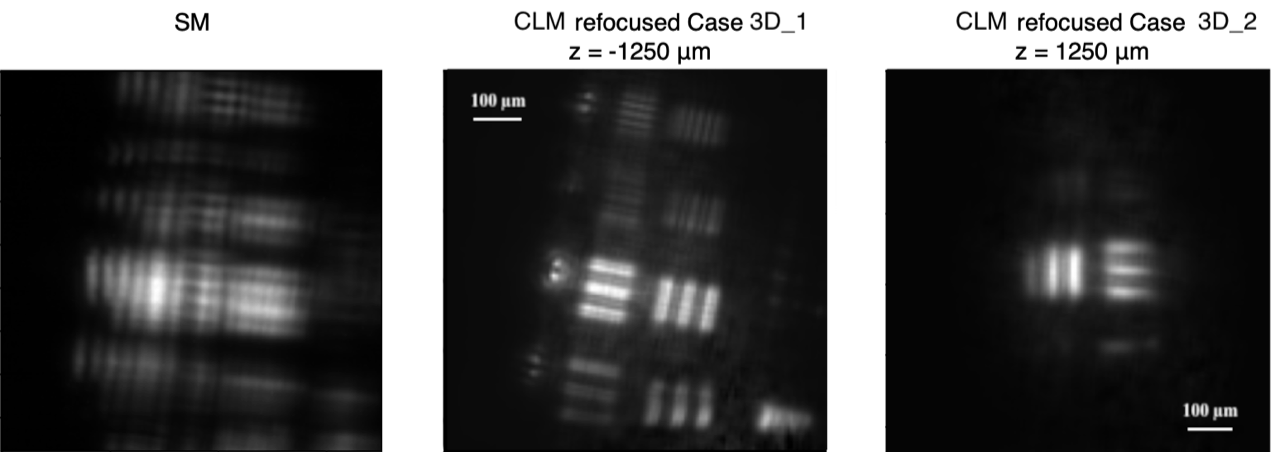}}
\caption{\textit{Panel} (a): Comparison of the resolution versus depth of field compromise in a standard microscope (SM) and in CLM with a numerical aperture of $0.23$ and illumination wavelength of $532\ \mathrm{nm}$. The curves represent the limit at which two slits with center-to-center distance $d$ (indicated on the vertical axis and equal to twice their width) can be discriminated at 10\% visibility, as a function of the longitudinal displacement from the objective focal plane ($f-f_O$). The curves are obtained assuming the same illumination wavelength and numerical aperture for the two microscopes. Points $3D_1$ and $3D_2$ indicate the parameters of the three-dimensional sample we have imaged (see lower panels), which is made of two planar resolution targets (with $d=49.6\,\mu\mathrm{m}$) placed at a distance $ f_1-f_O = -1250\,\mu\mathrm{m}$ and $ f_2-f_O = 1250\,\mu\mathrm{m}$, respectively, from the objective. Points A to E, and A' to D' correspond to further experimental data demonstrating the maximum achievable DOF of CLM at different resolutions, as the object (a resolution test target) is moved away from the plane at focus (A, A': $44.2\,\mu\mathrm{m}$ at $\pm 1000\,\mu\mathrm{m}$; B, B': $39.4\,\mu\mathrm{m}$ at $\pm 750\,\mu\mathrm{m}$; C, C': $31.3\,\mu\mathrm{m}$ at $\pm 500\,\mu\mathrm{m}$; D, D': $22.1\,\mu\mathrm{m}$ at $\pm 250\,\mu\mathrm{m}$, E: $4\,\mu\mathrm{m}$ at $f-f_o=0$). The corresponding images are reported in the Supplementary material. \textit{Lower panels}: images of the three-dimensional test sample described above, corresponding to points $3D_1$ and $3D_2$ in panel (a). \textit{Panel} (b) contains the image acquired by the standard microscope (SM): both targets are unresolved, as they are out of the DOF centered on the objective focal plane. \textit{Panels} (c) and (d) show the CLM refocused images of the closest ($3D_1$) and farthest ($3D_2$) planes of the three-dimensional test sample, respectively, as obtained from the same data employed in the left image of panel (b) by exploiting the additional directional information characterizing CLM to perform refocusing by means of Eq.~\eqref{refocus}.} 
\label{fig:triple_slit}
\end{figure}

\hspace{1 cm}

\noindent EXTENDED VOLUMETRIC IMAGING BY CLM

\hspace{1 cm}

\noindent The refocusing and depth mapping capability of CLM has preliminarily been tested with a simple and controllable three-dimensional object, made of two planar resolution targets (named $3D_1$ and $3D_2$) placed at two different distances from the objective lens. Correlation measurements combined with Eq.\ \eqref{refocus} have been employed to refocus the two test targets, separately, starting from data acquired by focusing the CLM to the focal plane of the objective lens, where neither one of the test targets was placed. In Fig.\ \ref{fig:triple_slit}, we report the results obtained by illuminating the same element on both targets, namely, triple slits with center-to-center distance $d = 49.6\,\mu\mathrm{m}$ and slit width $a=d/2$; the overall field of view (FOV) is $0.54\,\mathrm{mm}$. The test targets are placed at a distance of $2.5\, \mathrm{mm}$ from each other (\textit{i.e.} $f_{\text{3D}_1} - f_O = -1250\,\mathrm{\mu m}$ and $f_{\text{3D}_2}-f_O = 1250\,\mathrm{\mu m}$, where $f_O$ is the focal length of the objective lens), which is 
6 times larger than the natural depth of field of the microscope, at the given sample detail size (see Table~\ref{tab:comp}). Panel (b) reports the image acquired by the conventional microscope, which is clearly out of focus: none of the two triple slit objects can be recognized. However, as shown in panels (c) and (d), CLM enables refocusing both test targets. The reported results have been obtained by acquiring $N=5\cdot 10^3$ frames; we report in the Supplementary material further details on SNR characterization based on the number of acquired frames.

The improvement of CLM over standard microscopy is quantified in Fig.~\ref{fig:triple_slit}(a), where we report the resolution-versus-DOF compromise in the two microscopes. In particular, the curves indicate the limits at which two slits, of center-to-center distance $d$ equal to twice their width, can be discriminated at 10\% visibility, as a function of both the slit separation $d$ (on the vertical axis) and the distance from the objective focal plane (horizontal axis). The red points $3D_1$ and $3D_2$ indicate the experimental conditions corresponding to the data reported in Fig.~\ref{fig:triple_slit}(b). The plot indicates that CLM enables achieving a 6 times larger depth of field than standard microscopy, at the given resolution, or, alternatively, a 6 time better resolution, at the given depth of field. 
Remarkably, the sets of triple slits placed at $f_{\text{3D}_2}-f_O=1250\,\mu\mathrm{m}$ (i.e., the object placed farthest from the objective) are still perfectly resolved. This means that the resolution achieved on farther planes is not heavily influenced by the presence of other details placed along the optical path, despite the substantial spatial filtering that they perform.
The gray points labeled as A to E in Fig.~\ref{fig:triple_slit}(a) further demonstrate the agreement between the theoretical and the experimental combination of resolution and maximum achievable DOF of CLM. The range between $-1$ mm and $+1$ mm along the optical axis has been explored, in steps of $250\ \mu$m, by employing different triple-slit masks of the same planar resolution test target characterized by center-to-center distances ranging from $44\, \mu\mathrm{m}$ (A) to $4\, \mu\mathrm{m}$ (E). In particular, point E is close to the diffraction limit of a standard microscope and shows that CLM is capable of the same resolution at focus. All corresponding acquired CLM images are reported in the Supplementary material.

The results in Fig.~\ref{fig:triple_slit}(b)-(d) also demonstrate that CLM improves by over one order of magnitude the acquisition speed with respect to previous correlation-based light-field imaging protocols \cite{cpi_exp,cpi_snr}, where $5\times 10^4$ frames (against the current $5\times 10^3$) and additional low-pass Gaussian filtering were employed to achieve a comparable SNR. 
This improvement directly comes from the elimination of ghost imaging from the CLM architecture, and its replacement by conventional imaging at both sensor arrays. In fact, correlation between direct images has been shown to enable a significant improvement of the signal-to-noise ratio with respect to ghost imaging \cite{cpi_snr}. 

\begin{figure}
\centering
\subfigure[]{\includegraphics[width=0.30\textwidth]{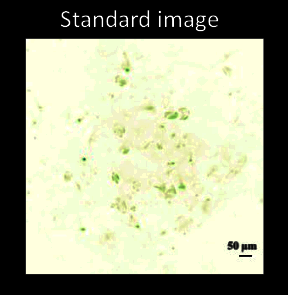}}
\\
\subfigure[]{\includegraphics[width=0.47\textwidth]{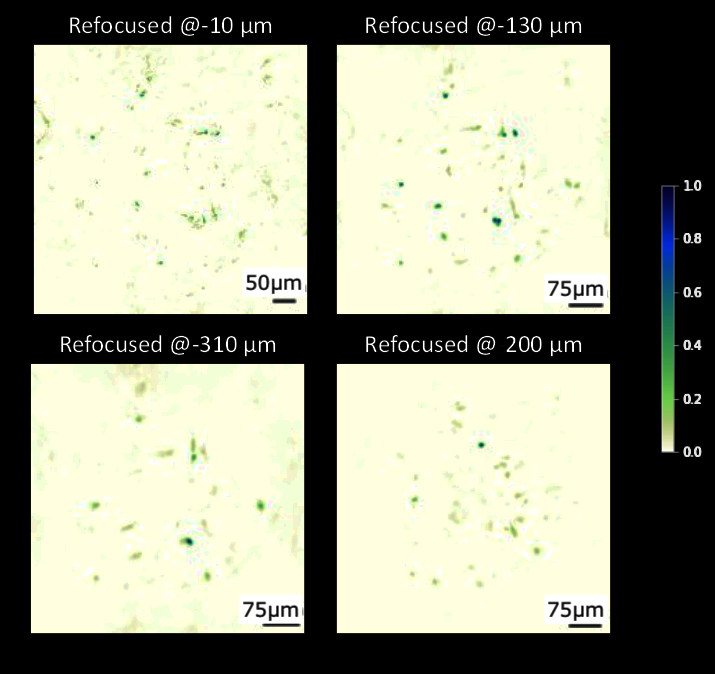}}
\caption{First-order image of a thick biomedical phantom (a) acquired at $f-f_O=0$, and refocused CLM images of four distinct planes inside the same sample (b); the specific value of $f - f_O$ is reported on top of each image. All images have been obtained from the same data, using $N=5000$ frames. Differential Correlation Plenoptic Imaging has been employed to remove the background due to light coming from out-of-focus planes (see Supplementary Material for details).}
\label{fig:amido}
\end{figure}

\begin{figure}
\centering
\includegraphics[width=0.5\textwidth]{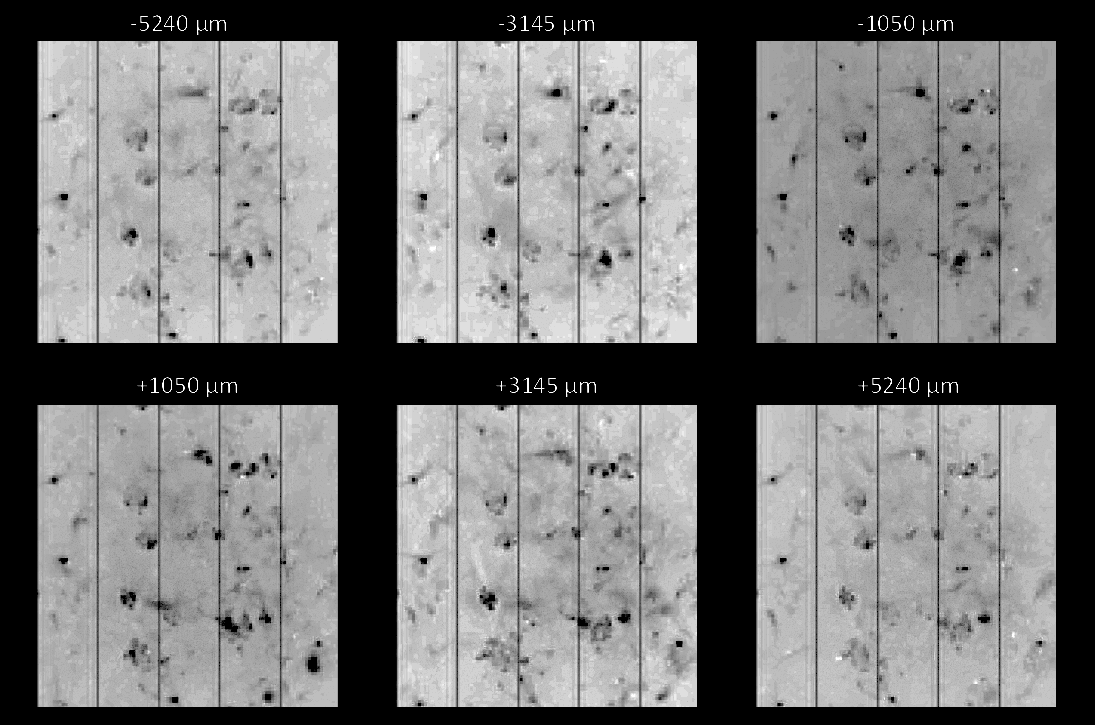}
\caption{Change of perspective within the $1\, mm^3$ sample of starch grains suspended in gel, as obtained through CLM. The change of perspective is implemented by correlating the intensity registered by each pixel of the spatial sensor ($D_a$) with the intensity registered by a small portion ($10 \times 10$ pixels) of the angular sensors ($D_b$) [i.e., by varying the second argument in Eq.~\eqref{Gammacorr}]. The selected portion along $D_b$ corresponds to a portion of the objective lens from which the sample is viewed. Here, we show a selection of six different points of view on the horizontal diameter of the objective lens, spanning a range of about $1.5$ cm. In the first sector on the left, a focused grain is observed (its position does not change with the perspective). In all other sectors, out-of-focus grains are seen to move when changing perspective.}
\label{fig:amido_prosp}
\end{figure}

After testing CLM with a binary, planar and known sample, we imaged a more complex, thick, biomedical phantom, made of birefringent starch granules ranging from $10$ to $40\,\mu\mathrm{m}$ dispersed in a transparent gel, within a FOV of $1\,\mathrm{mm}$ and a thickness of about $1\,\mathrm{mm}$. We performed birefringent imaging by placing the sample between crossed polarizers. 
The focused plane inside the sample was arbitrarily chosen at approximately half of the its thickness. In Fig.\ \ref{fig:amido}(a), we show the image of the focused plane, while Fig.\ \ref{fig:amido}(b) reports the refocused images of four different planes, longitudinally displaced from the focal plane by an optical distance of $-10\,\mu\mathrm{m}$, $-130\,\mu\mathrm{m}$, $-310\,\mu\mathrm{m}$, and $+200\,\mu\mathrm{m}$, respectively. It is evident that some aggregates appear focused in only one of the four images, which provide a tool to identify their longitudinal optical distance from the focal plane. The high volumetric resolution enabled by CLM is also demonstrated by refocusing, by means of the same acquired data, 41 planes over $1.4\,\mathrm{mm}$, with a longitudinal resolution of $30\,\mu\mathrm{m}$, all characterized by a FOV of $1\,\mathrm{mm}$ at a transverse resolution $\leq 20\,\mu\mathrm{m}$ (see video in the Supplementary Material); the voxel size is diffraction-limited, in line with expectations reported in Table I. This results comes from the capability of CLM to acquire, through a single correlation image, $130,000$ points-of-view of the sample, distributed over an area of $1.5\,\mathrm{cm}^2$ (on the objective plane), each one characterized by a diffraction-limited spatial resolution of $40 \,\mathrm{\mu m}$. 
In a standard light-field microscope, the same multiplicity of points of view would have led to a reduction of the resolution at least by a factor $360$, making the granules undetectable. In fact, in conventional lighfield microscopy, a maximum of $14\times 14$ perspectives are typically chosen, considering the associated reduced resolution (e.g., from $4000\times 4000$ to $296\times 296$) \cite{ng}. Interestingly, in the current CLM architecture, the SNR is high enough for images from different viewpoints to be effectively observable - hence available for further data analysis, such as three-dimensional reconstruction. 
The effect of the change of perspective obtained when moving (on the objective lens) along the horizontal direction is shown in Fig. \ref{fig:amido_prosp}: While the position of details at focus does not change with the particular perspective, out-of-focus starch granules shift along the horizontal direction as the point of view is changed.

\section{Discussion}

The refocusing capability of Correlation light-field microscopy has brought to a 6 times larger depth of field than in a conventional microscope having the same NA, and the same (diffraction-limited) resolution in the focused plane. These results are in excellent agreement with the expected refocusing range of CLM at the given resolution ($d=50\,\mu\mathrm{m}$), thus showing the reliability of the proposed CLM architecture. The volumetric resolution ($20 \times 20 \times 30 \,\mu\mathrm{m}^3$) achieved by CLM to image a complex thick sample ($1 \times 1 \times 1 \,\mathrm{mm}^3$) is a further very interesting result, considering the scanning-free nature of CLM. In this context, CLM has also been employed to recover about $130,000$ diffraction-limited perspectives over the three-dimensional sample. These results are unattainable by standard light-field microscopy, due to its hyperbolic tradeoff between spatial resolution and multi-perspective views (hence, maximum achievable DOF). 

In general, increasing the acquisition speed of CLM is the principal challenge that needs to be addressed to guarantee its competitivity with state-of-the art light-field microscopes \cite{microscopy4}. Such speed-up is in fact of paramount importance both for avoiding radiation damage of biomedical samples, for \textit{in vivo} imaging, and for studying dynamic processes.
The large SNR of CLM with respect to the original CPI scheme represents a first significant step in this direction, as it enabled to increase the acquisition speed by one order of magnitude, while still guaranteeing an even higher SNR (see Ref.~\cite{cpi_prl}). Similar to the Differential Correlation Plenoptic Imaging (DCPI) approach implemented here (Figs.~\ref{fig:amido}-\ref{fig:amido_prosp}), and outlined in the Materials and Methods section, a further step toward acquisition speed-up is compressive sensing and deep learning techniques, as increasingly applied to imaging tasks \cite{compressive1,compressive2,compressive3,deeplearning}. From the hardware viewpoint, the acquisition speed of our CLM has ample room for improving, both by investigating possible optimizations in our current acquisition routine and by employing cameras with better time performance. The most immediate way to start boosting the time performance of the current CLM, for example, is to employ the camera (see Sec. \ref{sec: materials}) in rolling-shutter mode, rather than global shutter, which we have been using to guarantee that the retrieved intensity patterns $I_a$ and $I_b$ (which are then correlated pixel by pixel) are  simultaneous statistical sampling of the chaotic source. This condition is consistent with the theoretical model (i.e., Eq.~\eqref{Gammacorr}), but it is certainly interesting to search for a regime in which moving slightly away from the theory introduces small enough artifacts to justify the gain in speed. With our camera, this could mean even doubling the frame rate, reducing current the acquisition time to about $20$ seconds (from the current $43$).
Also the chaotic source can be significantly improved by replacing the current ground-glass disk (see Material and Methods) with a digital micromirror device (DMD), which adds versatility and available statistics, while significantly decreasing the source coherence time due to its typical frame rate of about $30$ kHz. 
Also, since the DMD patterns are completely user-controllable, their features can be customized to achieve the desired SNR with the lowest number of frames possible, even experimenting with structured illumination. 
In this scenario, the acquisition speed will essentially be limited by the maximum frame rate of the sensor and, eventually, by the data transferring speed. 
This issue can be addressed by replacing our current sCMOS with faster cameras, capable of reaching $6.6$ kfps at full resolution \cite{camera}, or with ultra-fast high-resolution SPAD arrays, enabling acquisition rates as high as $10^5$ binary frames per second, in a $512 \times 512$ array \cite{epfl1,epfl2}. When choosing alternative cameras, speed should not be favored at the expenses of readout noise, dynamic range, detection efficiency, or minimum exposure time, all which are relevant parameters in correlation-based imaging. In this respect, SPAD arrays are of particular interest due to their much shorter minimum exposure time, ranging from a few hundreds of ps to 10 ns \cite{epfl1,epfl2,spadmi1,spadmi2}, although their binary nature may pose challenges. This would open up the possibility of replacing our controlled chaotic source with uncontrolled thermal sources, such as LEDs, lamps, and fluorescent samples, while preserving our capability to adequately measure intensity fluctuation correlations \cite{sunlight1,sunlight2}.

\section{Material and Methods} \label{sec: materials}

\hspace{1cm}

\noindent EXPERIMENTAL SETUP

\hspace{1cm}

\begin{figure}
\centering
\includegraphics[width=0.48\textwidth]{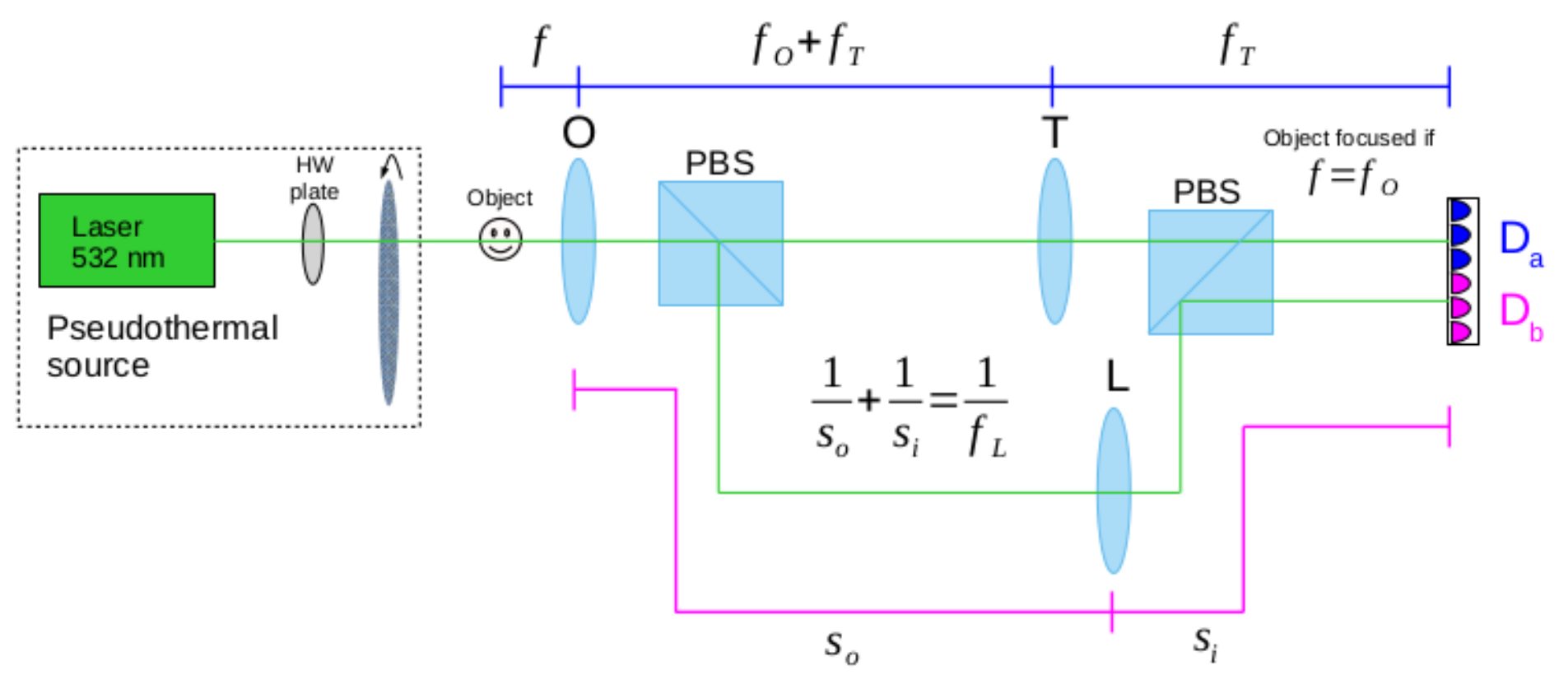}
\caption{Experimental setup for Correlation Plenoptic Microscopy. Light produced by a chaotic source, made from a laser, a half-wave plate and a rotating ground-glass disk, illuminates the object, is collected by the objective lens O, and reaches the first polarizing beam splitter (PBS). The transmitted beam passes trough the tube lens T and the second PBS, and finally impinges on the detector $\mathrm{D}_a$, which is the (blue) half of the sensor of the CMOS camera. The focused image is characterized by magnification $M = f_T/f_O = 4.2$.
The reflected beam passes through the additional lens L and then is reflected by a second PBS toward detector $\mathrm{D}_b$, which is the disjoint (magenta) half of the sensor on the same sCMOS camera. The magnification of the objective lens O on $\mathrm{D}_b$ is $M_L=0.31$.}\label{fig:schema_setup}
\end{figure}

\noindent The experimental setup employed to demonstrate CLM is shown in Fig.\ \ref{fig:schema_setup}. The controllable chaotic light source is a single-mode laser with wavelength $\lambda=532\,\mathrm{nm}$ (CNI MLL-III-$532$-$300$mW) illuminating a rotating ground glass disk (GGD), with diffusion angle $ \theta_d \simeq 14^{\circ} $, whose speed defines the source coherence time ($\approx 90 \mu s$). 
The laser spot size on the disk is enlarged to a diameter of 8 mm by a $6 \times$ beam expander, and the sample is placed at a distance of $10\,\mathrm{mm}$ after the GGD; the effective numerical aperture of our systems is thus $\mathrm{NA} = 0.23$  
which defines our expected diffraction-limited resolution $\delta = 1.6\,\mu\mathrm{m}$. 
Light transmitted by the object  propagates toward the objective lens O, with focal length $f_O = 30\,\mathrm{mm}$, and reaches the first polarizing beam splitter (PBS) where it is divided in two beams. The transmitted beam reaches the tube lens T, with focal length $f_T = 125\,\mathrm{mm}$, and then impinges on the part of the sensor identified with $\mathrm{D}_a$. The distance between the objective lens O and the tube lens T is equal to the sum of the focal lengths of the two lenses, $f_O + f_T$, and the distance between T and $\mathrm{D}_a$ coincides with $f_T$. The focused image plane thus lies at a distance $f_O$ from the objective lens. The beam reflected off the PBS illuminates lens L, with focal length $f_L = 150\,\mathrm{mm}$, then impinges on the part of the sensor identified with $\mathrm{D}_b$, after being reflected by the second PBS. The distance $S_O$ between the objective lens O and the lens L, and the distance $S_I$ between L and $\mathrm{D}_b$ are conjugated and the front aperture of the objective is imaged on $\mathrm{D}_b$. The measured magnification of such image is $M_L = 0.31$. Two disjoint halves of the same camera (Andor Zyla 5.5 sCMOS) are employed to simulate the two sensors $\mathrm{D}_a$ and $\mathrm{D}_b$, in order to guarantee synchronization. To fully exploit the dynamic range of the camera and maximize the SNR, we balance the intensities of the beams on the two halves of the sCMOS camera by means of a half-wave plate placed before in the laser beam, before the GGD. The camera sensor is characterized by $2560\times 2160$ pixels of size $\delta_{\mathrm{p}} = 6.5\,\mu\mathrm{m}$ and can work at up to 50 fps in full-frame mode (in global shutter mode, 100 fps with rolling shutter). Since the resolution $\delta = 1.6\,\mu\mathrm{m}$ on the object corresponds to a magnified resolution cell $M\delta = 6.7\,\mu\mathrm{m}$ on the sensor, data reported in Figs. \ref{fig:amido}, \ref{fig:amido_prosp} were generally acquired with a $2\times 2$ hardware binning; no binning was applied when acquiring data corresponding to points C, D, their primed counterparts, and E, in Fig. \ref{refocus}(a). 
The test targets employed to acquire data reported in Fig. \ref{refocus} are Thorlabs R3L3S1N and R1DS1N.

The exposure time was set at $\tau = 92.3\,\mu\mathrm{s}$ to match the coherence time fo the source, and the acquisition rate of the camera to $R = 120\,\mathrm{Hz}$, the maximum speed possible at our FOV in global shutter mode.

\hspace{1cm}

\noindent CLM WITH DIFFERENTIAL CORRELATION PLENOPTIC IMAGING APPROACH

\hspace{1cm}

Some of the images shown in this work (Figs. \ref{fig:amido}-\ref{fig:amido_prosp}) have been obtained by applying the refocusing algorithm \eqref{refocus} to a modified version of the correlation function reported in Eq.~\eqref{Gammacorr}, that was developed to enable noise reduction. The new correlation function reads
\begin{equation}
\widetilde{\Gamma}(\bm{\rho}_a,\bm{\rho}_b) = \langle \Delta I_a (\br_a) \left( \Delta I_b (\br_b) - K(\br_a,\br_b) \Delta I_a^{\mathrm{TOT}} \right) \rangle,
\end{equation}
with $I_a^{\mathrm{TOT}}$ the total intensity impinging on the detector $\mathrm{D}_a$, and
\begin{equation}
K(\br_a,\br_b) = \frac{\langle \Delta I_b (\br_b)\Delta I_a^{\mathrm{TOT}} \rangle}{\langle (\Delta I_a^{\mathrm{TOT}})^2\rangle},
\end{equation}
fixed by the condition of minimizing the variance of $\widetilde{\Gamma}(\bm{\rho}_a,\bm{\rho}_b)$, which is proportional to
\begin{align}
\mathcal{F}(\br_a,\br_b) = &\langle \left[\Delta I_a (\br_a) \left( \Delta I_b (\br_b) - K(\br_a,\br_b) \Delta I_a^{\mathrm{TOT}} \right) \right]^2 \rangle \nonumber \\ & - \widetilde{\Gamma}(\bm{\rho}_a,\bm{\rho}_b)^2 .
\end{align}
The approach employed to estimate this new correlation function, inspired by differential ghost imaging \cite{gatti_DGI}, has been named \textit{Differential Correlation Plenoptic Imaging}. A detailed treatment will be given elsewhere.

\hspace{1cm}

\noindent VIEWPOINT MULTIPLICITY

\hspace{1cm}

\noindent The viewpoint multiplicity is defined as the number of viewpoints per transverse direction. In the case of CLM, the viewpoint multiplicity is estimated as the number of resolution cells falling within the diameter $D$ of the objective lens. The resolution cell must be evaluated by considering that, in the correlation function, the sample acts as an aperture, and thus determines the resolution on the objective lens (see Ref.~\cite{cpm_theory} for details). Considering a sample made of a bright object of diameter $a$, placed in the focal plane of the objective lens, the size of the resolution cell on the lens plane reads
\begin{equation}
\Delta x_{\mathrm{lens}} = 1.22\, \frac{\lambda f_O}{a}.
\end{equation} 
Therefore, the viewpoint multiplicity can be evaluated as
\begin{equation}
\frac{D/2}{\Delta x_{\mathrm{lens}}} = \frac{D}{1.22\, \lambda f_O} a = \frac{a}{\Delta x_0}.
\end{equation}
This result highlights an interesting reciprocity relation between the two apertures and the two resolution cells, on the objective lens and the object plane.

\bigskip

\section*{Acknowledgments}
The Authors thank Francesco Scattarella for making the videos reported in the Supplementary material. This work was supported by Istituto Nazionale di Fisica Nucleare (INFN) project PICS, PICS4ME, TOPMICRO, and, partially, ``Qu3D'', and by Ministero dell'Universit\`a e della Ricerca (MUR) PON ARS project ``CLOSE -- Close to Earth''. Project Qu3D is supported by the Italian Istituto Nazionale di Fisica Nucleare, the Swiss National Science Foundation (grant 20QT21$\_$187716 ``Quantum 3D Imaging at high speed and high resolution''), the Greek General Secretariat for Research and Technology, the Czech Ministry of Education, Youth and Sports, under the QuantERA programme, which has received funding from the European Union's Horizon 2020 research and innovation programme.

\section*{Author Information}

\subsection*{Contributions}
M.D. and F.V.P. conceived the idea and wrote the paper. M.D., A.G. and F.V.P. supervised the work. A.S. contributed to the development of the theory and the algorithms, to the first design of the experimental setup, and, together with F.D., to the preliminary data acquisition. F.D. has contributed to the improvement of the setup and to the optimization of the data analysis program. G.M. has further optimized the data analysis program, developed and implemented the DCPI approach, contributed to writing the paper, and, in collaboration with D.G., performed the experiment. G.S. provided support for the design of the experiment, and contributed to writing the paper. All authors has contributed to revising the manuscript, and approved its final version.

\subsection*{Corresponding author}
Correspondence to Milena D'Angelo (milena.dangelo@uniba.it) and Francesco V. Pepe (francesco.pepe@ba.infn.it).

\onecolumngrid

\appendix

\section{Supplementary Material}

\begin{figure}
\centering
\includegraphics[width=0.9\textwidth]{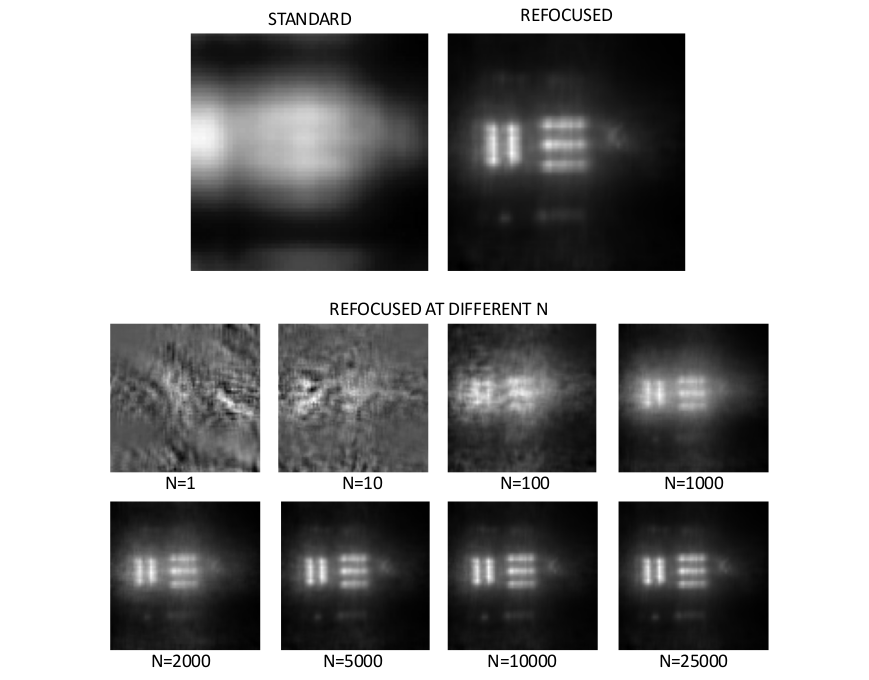}
\caption{\textit{Upper panels.} Comparison between images of a triple slit ($d=44.2\,\mu\mathrm{m}$) placed out of focus ($ f-f_O = 1.00\mathrm{mm}$), as directly acquired by the microscope (left panel) and refocused by CLM (right panel). \textit{Middle and lower panels.} Images of the same triple slit refocused by using a different number $N$ of frames: middle panel, from left to right, $N=1,10,100,1000$, lower panel, from left to right: $N=2000,5000,10000,25000$.}\label{fig:n_frames}
\end{figure}

Figure \ref{fig:n_frames} reports four cases of refocused images of a triple slit ($d=44.2\,\mu\mathrm{m}$, corresponding to point A' in Fig. 3(a) of the main text) placed outside the DOF of the conventional microscope ($ f-f_O = 1.00\,\mathrm{mm}$); CLM refocusing has been implemented by computing correlations within different numbers of acquired frames $N$, ranging from $1$ to $25,000$. The image quality clearly improves as the number of frames increases from $100$, where the image of the object begins to be observable, to $5\times 10^3$, while improvements become negligible for larger number of frames. Since the sample is a binary object (\textit{i.e.} made up of either transmissive or non-transmissive details), an estimation of the final image quality can be obtained by considering a signal-to-background ratio (SBR), where the image of transmissive parts of the object is considered as signal, and the remaining parts of the frame as background. We define the SBR by normalizing the refocused images from $0$ to $1$ and separating pixels in the two subsets
\begin{align}
I_{\mathrm{signal}} & = \{ (i,j) \text{ such that } \Sigma_{i,j} > 1/2 \} , \\
I_{\mathrm{background}} & = \{ (i,j) \text{ such that } \Sigma_{i,j} \leq 1/2 \} ,
\end{align}
with $\Sigma_{i,j}$ the normalized signal and $(i,j)$ the indeces identifying pixel positions. The cardinalities $|I_{\mathrm{signal}}|$ and $|I_{\mathrm{background}}|$ are then used to compute the mean values of the signals in the two sets of data:
\begin{align}
\overline{\Sigma}^{\mathrm{(signal)}} & = \frac{1}{|I_{\mathrm{signal}}|} \sum_{(i,j)\in I_{\mathrm{signal}}} \Sigma_{i,j}  , \\
\overline{\Sigma}^{\mathrm{(background)}} & = \frac{1}{|I_{\mathrm{background}}|} \sum_{(i,j)\in I_{\mathrm{background}}} \Sigma_{i,j} , 
\end{align}
from which we get:
\begin{equation}\label{SBR}
\mathrm{SBR} = \frac{\overline{\Sigma}^{\mathrm{(signal)}}}{\overline{\Sigma}^{\mathrm{(background)}}}.
\end{equation}
The values of the SBR characterizing the CLM refocused images of Fig.\ \ref{fig:n_frames}, are reported in Table~\ref{tab:comp} for a number $N$ of acquired frames ranging from $1$ to $25,000$. Interestingly, the triple slit is already visible with $N = 100$, with a SBR $= 3.2$; this indicates an improvement of the SBR by one order of magnitude in the present scheme with respect to the original CPI scheme (see Ref. [23] of the main text). Though the theoretical expectation for true chaotic illumination supports a $\sqrt{N}$ scaling of the quantity reported in Table~\ref{tab:snr}, we observe an evident saturation effect after $N=5,000$ frames. This is related with the fact that our chaotic source is based on a rotating ground-glass disk, hence, the patterns of chaotic light that illuminates the transmissive parts of the test target, in each frame, are available in a finite number and are not completely independent from each other.

\begin{table}
\centering
\begin{tabular}{cc}
$N$ & SBR \\ 
\hline \hline 
1 & 1.0 \\  
10 & 1.7 \\  
100 & 3.2 \\
1,000 & 4.2 \\
2,000 & 4.6 \\
5,000 & 5.7 \\
10,000 & 5.8 \\
25,000 & 6.2 \\
\hline
\end{tabular}
\caption{Signal-to-background ratios (SBR) of the CLM refocused images of the triple slit masks reported in Fig.\ \ref{fig:n_frames}, computed from Eq.\ \eqref{SBR} by employing a varying number $N$ of acquired frames.}\label{tab:snr}
\end{table}

\begin{figure}[t!]
\centering
\subfigure[]{\includegraphics[width=0.60\textwidth]{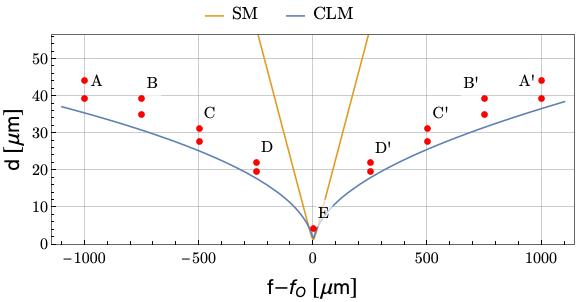}}
\\
\subfigure[]{\includegraphics[width=0.99\textwidth]{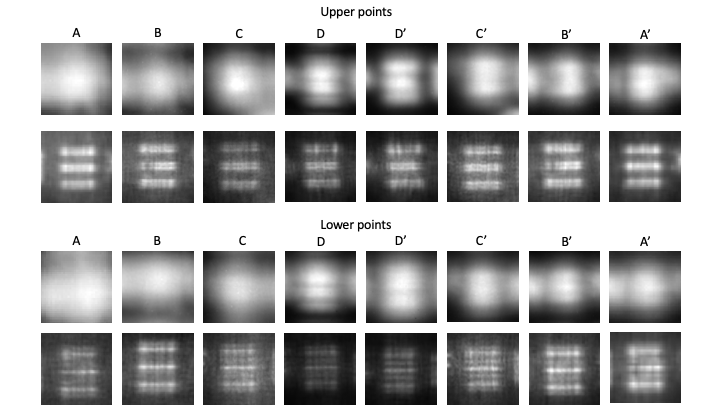}}
\\
\subfigure[]{\begin{tabular}{c||cccc||cccc||}
		& A & B & C & D & D' & C' & B' & A' \\
		Displacements $(f-f_O)$ [$\mu\,$m] & -1000 & -750 & -500 & -250 & 250 & 500 & 750 & 1000 \\
		\hline\hline
		$(d)$ [$\mu$m] for upper points & 44.2 & 39.4  & 31.3 & 22.1 & 22.1 & 31.3 & 39.4 & 44.2 \\
		$(d)$ [$\mu$m] for lower points & 39.4 & 35.1 & 27.8 & 19.7 & 19.7 & 27.8 & 35.1 & 39.4 \\
\end{tabular}}
\caption{Demonstration of the resolution versus DOF improvement of CLM, as the object (a resolution test target) is moved away from the focused plane. \textit{Panel} (a): 10\%-visibility limits in standard imaging (solid line, orange) and CLM (solid line, blue), as reported in Fig.~3(a) of the main text, as a function of the longitudinal displacement $f-f_O$ from the objective focal plane. The pairs of red dots labeled A to D represent the combinations of center-to-center slit distance $d$ and displacement $f-f_O$ at which the experimental data reported in panel (b) have been acquired, as detailed in table (c). Points labeled with a primed letter are placed symmetrically with respect to the focused plane to their non-primed counterpart. Point E is an acquisition with the target in the focused plane, and the corresponding acquired image is shown in Fig.~\ref{fig:focus+coherence} (left bottom panel). \textit{Panel} (b): the upper (lower) part of the panel shows the comparison between the unfocused micrsocope (SM) images and the CLM refocused images for the each element of the upper (lower) series of points indicated in panel (a). The table in \textit{panel} (c) reports the values of $f-f_O$ and $d$ related with the different employed samples.}
\label{fig:calibration}
\end{figure}

\begin{figure}
\includegraphics[width=0.5\textwidth]{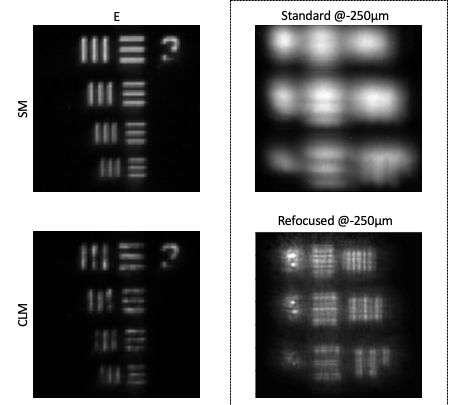}
\caption{\textit{Left}: comparison between the standard microscope (SM) and CLM images in the focused plane. The four smaller elements of group 7 of the resolution test target are imaged ($6.2\,\mu\mathrm{m}$, $5.5\,\mu\mathrm{m}$, $4.9\,\mu\mathrm{m}$, and $4.4\,\mu\mathrm{m}$) by CLM with the same resolution as in a conventional microscope. \textit{Right}: microscope and refocused image of elements 4, 5, 6 (from bottom to top) of group 5 of the test target ($22.1\,\mu\mathrm{m}$, $19.7\,\mu\mathrm{m}$, $17.5\,\mu\mathrm{m}$), placed out of focus at a longitudinal distance $f-f_O= - 250\,\mu\mathrm{m}$, which is very close to the 10\%-visibility curve of CLM. In the CLM image, as the separation between neihghboring slits decreases, fringes due to interference appear and the slits can no longer be resolved. The appearance of non-resolved details is very different from that typical of standard imaging (upper image) due to the intrinsic coherent nature of CLM. 
}
\label{fig:focus+coherence}
\end{figure}

Fig. \ref{fig:calibration} and \ref{fig:focus+coherence} demonstrate the improved performances of CLM over a standard microscope, in terms of resolution and depth of focus. Fig. \ref{fig:calibration} shows the accordance between the theoretical resolution, estimated as the distance between two slits that can be refocused with 10\% visibility, and the experimental one. In particular, we explore the range between $-1$ mm and $+1$ mm along the optical axis, in steps of $250\ \mu$m. The two solid lines in Fig.~\ref{fig:calibration} panel (a) are the 10\% visibility curves for standard microscopy (orange) and CLM (blue). Each pair of red dots, labeled as A to D, represent the center-to-center distance in triple-slit masks of a 1951 USAF resolution test target. Two resolutions have been tested for each different position of the target along the optical axis. At any given position on the optical axis, we employ the smallest element on the target having slit distance larger than the theoretical visibility curve (lower point in each pair) and the immediately larger element (upper point in the pair). Pairs labeled A' to D' are symmetrically displaced, with respect to the focused plane, to their non-primed counterparts. The comparison between the (unfocused) microscope images and the CLM refocused images is shown in Fig. \ref{fig:calibration} panel (b). Upper points in each pair are very clearly resolved, as prescribed by the theory. Points closer to the 10\% visibility (i.e., lower points), on the other hand, are less resolved and offer the chance to analyze the effect of the loss of resolution in CLM refocused images: All of them, in fact, show the typical fringes of coherent imaging. This is not unexpected, since CLM is by all means a coherent imaging technique, although based on an incoherent source of light [29]. The effect of the loss of resolution in CLM is also displayed on the right panel of Fig. \ref{fig:focus+coherence}, where elements 4, 5 and 6 of group 5 of the test target (corresponding to $22.1\ \mu m$, $19.7\ \mu m$ and $17.5\ \mu m$ resolutions) are visible. Element 4 is the lowest one in the image and corresponds to the upper point of pair D in Fig. \ref{fig:calibration}. As the details become smaller, as in element 5 (lower point of pair D), fringes appear between the slits and keep degrading the image resolution to the point where they prevent the three slits from being distinguished (element 6).
Point E in Fig.~\ref{fig:calibration} is close to diffraction limit of a standard microscope and shows that CLM is capable of the same resolution at focus of a conventional microscope with the same NA. However, since the optics and magnifications have been chosen having in mind imaging of samples with cell-like details (i.e., tens of microns), the diffraction-limited resolution of the microscope cannot be reached due to the size of the pixels; in fact, with our magnification of $4.2$ and pixel size of $6.5\,\mu\mathrm{m}$ (see Material and Methods), adjacent resolution cells ends up in the same pixel. The left side of Fig.~\ref{fig:focus+coherence} shows the comparison between the microscope and CLM images in the focused plane. The last four elements of group 7 of the test target are perfectly resolved to the minimum available slit distance of $4.4\,\mu\mathrm{m}$. Also, as one would expect from the visibility curves, the resolution is symmetric with respect to the focused plane.

\begin{figure}[t!]
\centering
\includegraphics[width=0.9\textwidth]{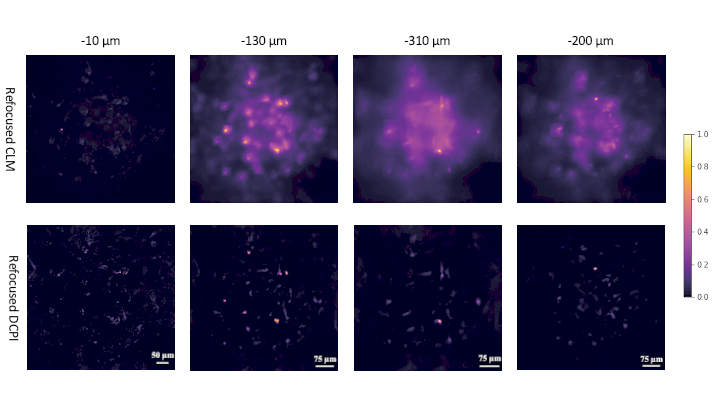}
\caption{CLM refocused images of four distinct planes inside a starch dispersion in gel (upper panels) and the corresponding images obtained by applying the DCPI algorithm (lower panels). All images were elaborated from the same acquisition of the correlation function using $N=5000$ acquired frames, as in the main text. Data in the bottom row are the same reported in Fig. 4 of the main text, but in a different color scale that makes background suppression more evident.}
\label{fig:clm_dclm}
\end{figure}

As mentioned in the Material and Methods section of the main article, we have developed a novel approach to data analysis in CPI, named Differential CPI (DCPI), that enables further improving its SNR performances (paper in preparation). Here we employ DCPI to get rid of the spurious background coming from out-of-focus planes within the refocused CLM images of complex three-dimensional samples. In Fig.\ \ref{fig:clm_dclm}, we compare the bare results obtained by refocusing of a complex thick three-dimensional sample (same data employed in the main text for the dispersion of starch in gel) with the one obtained by applying the DCPI algorithm directly on the refocused images. DCPI clearly improves the quality of the CLM refocused images by cleaning up the typical background due to out-of-focus planes within the three-dimensional sample.

\end{document}